\documentclass{PoS}

\usepackage[square,comma,numbers,sort&compress]{natbib} 
\usepackage[utf8]{inputenc}
\usepackage{graphicx}
\usepackage{color}
\usepackage{amsfonts}
\usepackage{amsmath}

\title{{\vspace{-18mm} \normalsize\hfill{\small DESY 16-207}}\\[10mm]
Finite Size Scaling of the Higgs-Yukawa Model near the Gaussian Fixed Point}

\ShortTitle{Finite Size Scaling of the Higgs-Yukawa Model near the Gaussian Fixed Point}

\author{\speaker{David Y.-J. Chu} \\
	  National Chiao-Tung University, Hsinchu, Taiwan\\
	  E-mail: \email{ren1072.ep99@g2.nctu.edu.tw}}

\author{Karl Jansen\\
	NIC, DESY Zeuthen, Germany\\
	E-mail: \email{karl.jansen@desy.de}}

\author{Bastian Knippschild\\
	HISKP, Bonn, Germany\\
	E-mail: \email{knippschild@hiskp.uni-bonn.de}}

\author{C.-J. David Lin\\
	National Chiao-Tung University, Hsinchu, Taiwan\\
	E-mail: \email{dlin@mail.nctu.edu.tw}}
       
\author{Attila Nagy\\
	Humboldt University Berlin; NIC, DESY Zeuthen, Germany\\
	E-mail: \email{nagy@physik.hu-berlin.de}}

\abstract{

We study the scaling properties of Higgs-Yukawa models.
Using the technique of Finite-Size Scaling, we are able to 
derive scaling functions that describe the observables of the model
in the vicinity of a Gaussian fixed point.
A feasibility study of our strategy is performed for the pure 
scalar theory in the weak-coupling regime.  
Choosing the on-shell renormalisation scheme gives us an advantage to fit 
the scaling functions against lattice data with only a small
number of fit parameters.
These formulae 
can be used to determine the universality of the observed 
phase transitions, and thus play an essential role in future
investigations of Higgs-Yukawa models, in particular in the strong Yukawa
coupling region.

}

\FullConference{34th annual International Symposium on Lattice Field Theory\\
                 24-30 July 2016\\
                 University of Southampton, UK}

\begin{document}

\section{Introduction}
 
The Standard Model (SM) is a very successful theory for explaining the 
interactions amongst elementary particles.  However the Higgs 
sector of the SM, represented by the $O(4)$ scalar model, is known to be trivial
\cite{Frohlich:1982tw, Luscher:1987ay, Bernreuther:1987hv, Korzec:2015pma}.  
This means that we cannot dispense the ultraviolet cutoff in this model else the renormalisation
group (RG) flow will render it non-interacting.  To be more specific,
RG flows always approach Gaussian fixed-points (GFP) in these models, where the couplings 
go to zero logarithmically as the ultraviolet cutoffs are raised to infinity.

The situation may differ when we consider the Higgs-Yukawa (HY) sector.
Recent works using perturbative analyses \cite{Molgaard:2014mqa, Hung:2009hy}
suggest the possible existence of non-trivial fixed point in the HY models.
This presence of non-trivial fixed points can lead to UV-completion.
Our recent publication \cite{Bulava:2012rb} using lattice study of a HY model 
also reported a phase transition at large bare yukawa couplings.
The critical exponents of the phase transition were extracted using finite-size scaling (FSS), and found 
to be about two standard-deviations away from the mean-field scaling law.
However we could not rule out the possibility for the fixed point being a GFP,
where the logarithmic behaviour may play an important role in our
FSS study \cite{Frohlich:1982tw, Kenna:2006hr}.

We extend the strategy a'la Brezin and Zinn-Justin \cite{Brezin:1985xx} 
to study the FSS of systems including fermions.  
To the best of our knowledge, this work is the first addressing the FSS in HY-like models.
The leading logarithmic scaling corrections are 
obtained using one-loop perturbation theory.
The feasibility test of our strategy is performed with the scalar $O(4)$ model.
Adopting the on-shell subtraction scheme allows us to test the lattice data with
only a small set of fit parameters for our scaling functions.

\section{Strategy}
 
We study FSS at hand of HY models containing $N$ degenerate scalars in the vicinity of its GFP.
Throughout this work, asymmetric lattices have been used with volume $L^3 \times T$ 
where $T=sL$ is the Euclidean time extent.  Quantities with caret are measured 
in lattice units.

\subsection{Finite-Size Scaling}

The universality class of a phase transition is characterised by the RG
running behaviour of the couplings near the fixed-points.  
The GFP in four dimensions is special as the beta function contains a double zero, 
leading to logarithmic corrections to the FSS \cite{Frohlich:1982tw}.  
Consider a bare correlator on a lattice, $X_0\left[ M^2_0, \lambda_0, y_0; a^{-1}, L \right]$, 
at zero external momentum, with classical dimension $D_X$.  It depends on the bare relevant
scalar quadratic coupling, $M_0^2$, bare scalar quartic coupling, $\lambda_0$, bare yukawa coupling, $y_0$,
cutoff, $a^{-1}$, and box length, $L$.  
The RG analysis leads to,
\begin{eqnarray}
\label{FSS_analysis}
 \hat{X}_0 \left[ M_0^2 , \lambda_0, y_0; a^{-1}, L \right] Z_{X} (a^{-1},m_P) 
 &=& \hat{X} \left[ M^2 (m_P), \lambda(m_P), y(m_P); m_P, L \right],  \nonumber \\
 &=& \zeta_{X} (m_PL) \hat{L}^{-D_{X}} \hat{X} 
 \left[ \hat{M}^2 (L^{-1}) \hat{L}^{2}, \lambda(L^{-1}), y(L^{-1}) \right], \\
 \text{where  } \zeta_{X} (m_PL) &=& \exp\left( \int_{m_P}^{L^{-1}} \gamma_X(\rho) \frac{d\rho}{\rho} \right) ,
\end{eqnarray}
where $Z_{X} (a^{-1},m_P)$ represent the relevant wavefunction renormalisation constants.  
In the first step of Eq.~(\ref{FSS_analysis}) we adopt 
on-shell subtraction scheme to renormalise this correlator at scale $m_P$,
which is the pole mass of the massive scalar propagator.  
We perform RG running to a renormalisation scale chosen as the 
inverse box length $L^{-1}$ in the following step by 
$\zeta_{X} (m_PL)$, where the $\gamma_X$ is the anomalous dimension of the correlator, $X$.
In the end we perform a rescaling with the box length $L$, such that the scalar quadratic
coupling is accompanied with $L^2$ to form a dimensionless combination. 

If the renormlised system is in the vicinity of a non-trivial fixed-point, the RG behaviour
leads to a power-law dependence on the volume.
If the fixed-point is a GFP, the RG behaviour leads  
to a logarithmic dependence on the volume instead.
In the latter case, the scaling functions can be derived using perturbation theory to leading-order
as detailed in \cite{Chu:2015jba}.  The scaling variable is,
\begin{equation}
\label{scaling_variable}
 z=\sqrt{s}\hat{M}^2 \left( L^{-1} \right) \hat{L}^2 \lambda \left( L^{-1} \right)^{-1/2},
\end{equation}
where the couplings, $\hat{M}^2 (L^{-1})$ and $\lambda (L^{-1})$, 
are renormalised at $L^{-1}$.  
The logarithmic dependence on the volume can be obtained from solving the RG equations 
for these two couplings \cite{Chu:2015jba}.
The scaling function of the $n-$th moment of the scalar zero mode at scale $L^{-1}$ is,
\begin{equation}
\label{n-th_moment}
 \langle \hat{\varphi}^n \rangle \left( L^{-1} \right)= \left[ sL^4\lambda \left( L^{-1} \right) \right]^{-n/4}\
 \left[ \frac{\bar{\varphi}_{N+n-1}(z)}{\bar{\varphi}_{N-1}(z)} \right],
\end{equation}
where we denote $\hat{\varphi}$ as the radial component of the scalar zero mode. 
The definition of the functions $\bar{\varphi}_{N}(z)$ can also be found in \cite{Chu:2015jba}.

\subsection{On-shell Subtraction Scheme}

The renormalised quadratic coupling, $M^2$, is a crucial piece in Eq.~(\ref{scaling_variable}).  
In order to solve for $M^2(L^{-1})$, we have to consider multiplicative and additive 
renormalisation, therefore more fit parameters appear in the scaling formulae.
We can avoid fitting these parameters by adopting the on-shell subtraction
scheme in Eq.~(\ref{FSS_analysis}) and obtaining $M^2$ using lattice data.

The on-shell subtraction scheme is defined as the pole mass, $m_P$, identified with 
the renormalised mass at the scale of $m_P$ itself.
However the pole mass on the lattice suffers finite-volume effect arising from the light 
modes wrapping around the world.  
We used simple linear extrapolation in $\hat{L}^{-2}$ to infinite volume 
at the same lattice spacing to remove this effect.
In principle the massive modes also contributes to the finite-volume effect 
in the form of $e^{ -\hat{m}_P\hat{L} }$ \cite{Luscher:1985dn}, 
we will investigate this in the near future.
In order to apply leading-order perturbation theory, we need to be 
very close to critical point while keeping $m_P$ much lower than the cut-off scale.
This leads to the requirement $\hat{m}_P \ll 1$ 
as well as $\hat{L} \gg 1$ but with $\hat{m}_P\hat{L} \gtrsim 1$ for our studies.

We note that the relation between $m_P$ and renormalised quadratic coupling is different in symmetric and broken phases.
A more precise identification is by considering a quartic effective potential at the scale of $m_P$ and 
identifying $m_P^2$ with the second derivative of the potential at the minimum.
Denoting symmetric phase by (sym.) and broken phase by (bro.), the relation in the two phases is,
\begin{eqnarray}
 M^2 (m^{sym.}_P) &=&  (m^{sym.}_P)^2 , \nonumber \\
 M^2 (m^{bro.}_P) &=& -\frac{1}{2} (m^{bro.}_P)^2.
\end{eqnarray}
Last, we stress that since the pole mass can be different in the two phases,
such that we fit the data differently in the two phases.

\section{Numeric Results}

In order to test our strategy for FSS, 
with formulae that become very complicated when fermions are added, 
we have chosen the scalar O(4) model for a first feasibility study. 

\subsection{O(4) Model}

We present preliminary results for examining the logarithmic
scaling of the $O(4)$ model.  The action of this model is,
\begin{equation}
\label{Action}
 S = \int d^4x \left\{ \frac{1}{2}\partial_{\mu} \Phi^{T} \partial^{\mu} \Phi + 
 \frac{1}{2} M^2_0 \Phi^{T}\Phi + \lambda_0 \left( \Phi^{T}\Phi \right)^2 \right\},
\end{equation}
where $\Phi = ( \phi_1, \phi_2, \phi_3, \phi_4)^{T}$.  We have performed simulations
from $\hat{L}=12$ to $\hat{L}=36$ in steps of $2$, and $\hat{L} = 40,44$, 
using the cluster algorithm \cite{wolff1989collective} combined with metropolis updates.
Simulations are performed on asymmetric lattices with fixed ratio $s=2$, fixed bare quartic coupling
$\lambda_0 = 0.15$, and by scanning in bare quadratic coupling $M^2_0$.

The one-loop results for the running of the couplings in this $O(4)$ model leads to,  
\begin{eqnarray}
\label{pure_scalar_quartic_RGE}
 \mu \frac{d \lambda (\mu)}{d \mu} &=& \frac{6}{\pi^2} \lambda (\mu)^2, \\
\label{pure_scalar_qudratic_RGE}
 \mu \frac{d \hat{M}^2 (\mu)}{d \mu} &=& \frac{3}{\pi^2} \lambda(\mu) \hat{M}^2 (\mu).
\end{eqnarray}
Integrating these two equations from the scale of $m_P$ to $L^{-1}$, as in Eq.~(\ref{FSS_analysis}),
we obtain the expressions for $\lambda(L^{-1})$ and $\hat{M}^{2}(L^{-1})$.
Substituting the two expressions in Eq.~(\ref{scaling_variable}) gives us the scaling variable 
for the $O(4)$ model,
\begin{equation}
\label{O4_scaling_variable}
 z = \sqrt{2} \hat{M}^2(m_P)\hat{L}^2 \lambda(m_P)^{-\frac{1}{2}},
\end{equation}
where the exponents of the logarithms cancel.  
Note that the two pole masses lead to two different scaling variables in the two phases.

\subsection{Fitting Results}

\begin{figure}[!t]
\vspace*{-0.18 cm}
\begin{center}
\includegraphics[width=0.5\columnwidth]{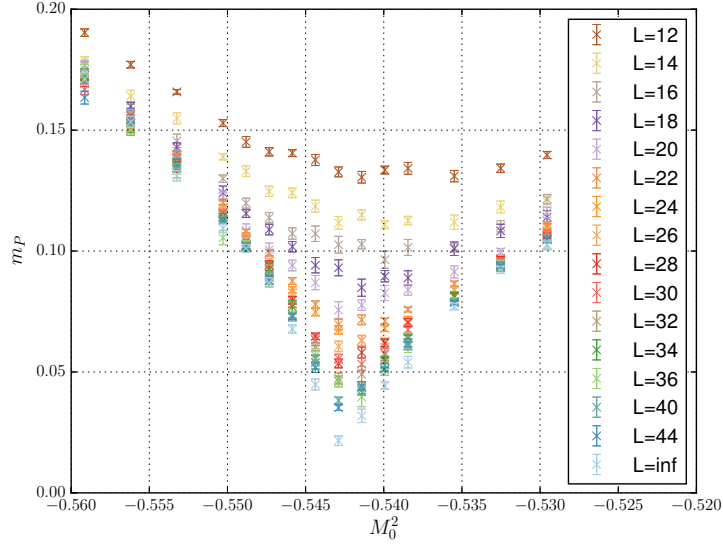}
\caption{
The pole masses at finite volumes and the results of infinite-volume extrapolations.
  \label{fig:extrapolation} 
}
\end{center}
\end{figure}

We have performed a study using our scaling formulae for the vacuum expectation value (VEV), the susceptibility,
and Binder's Cumulant, respectively.  Firstly, we present
infinite-volume extrapolation for removing the effects of light modes wrapping around the world.
The smallest volume we used in the extrapolation is chosen to be $\hat{L} = 28$ 
to meet the requirement $\hat{m}_P\hat{L} \gtrsim 1$ for all our data.
The results are shown in Fig.~\ref{fig:extrapolation}.  

We study Binder's Cumulant, $Q_L$.  Its functional form can be found using Eq.~(\ref{n-th_moment}),
\begin{figure}[!t]
\begin{center}
\vspace*{-0.2 cm}
\includegraphics[width=0.45\columnwidth]{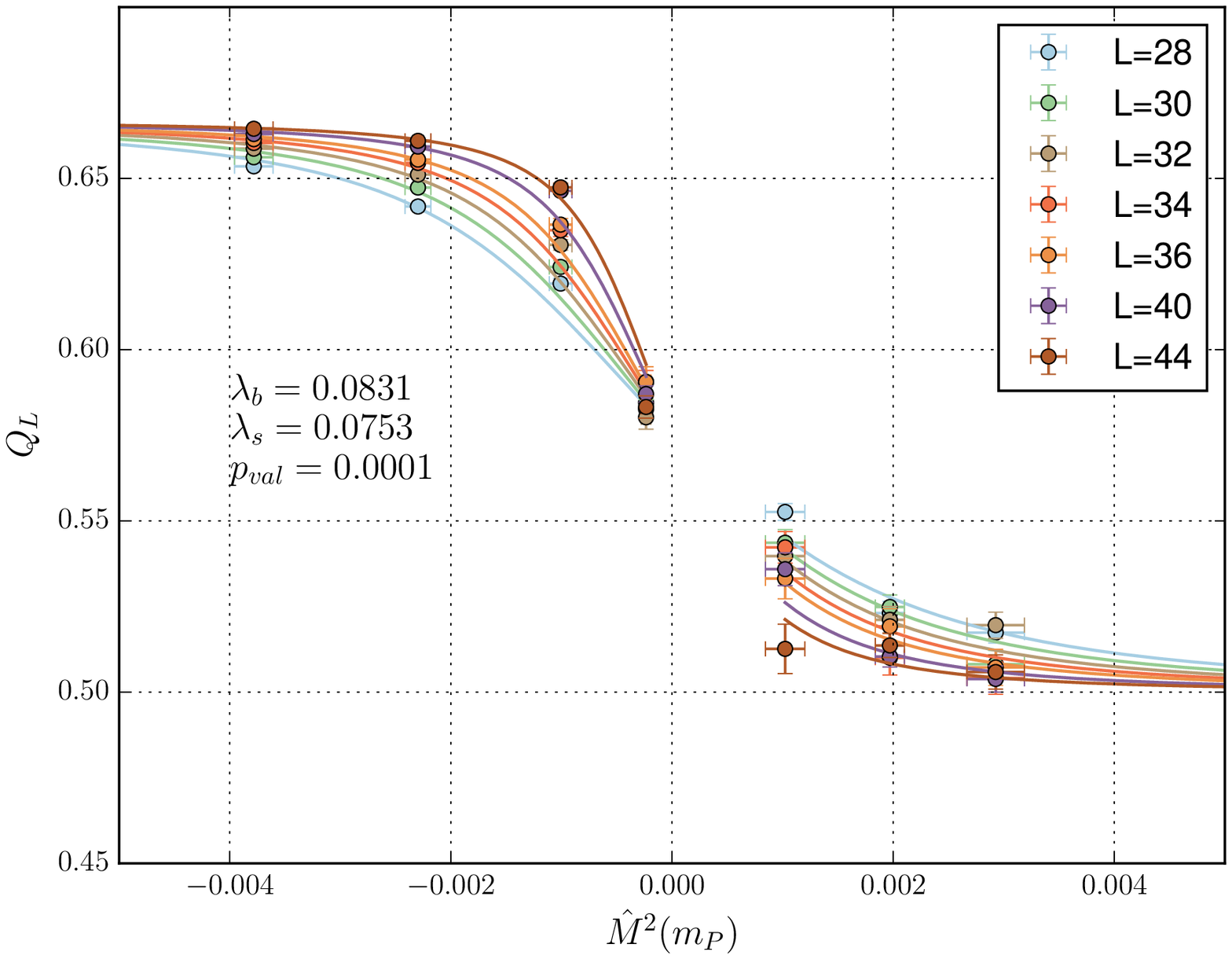}
\hspace*{13 mm}
\includegraphics[width=0.45\columnwidth]{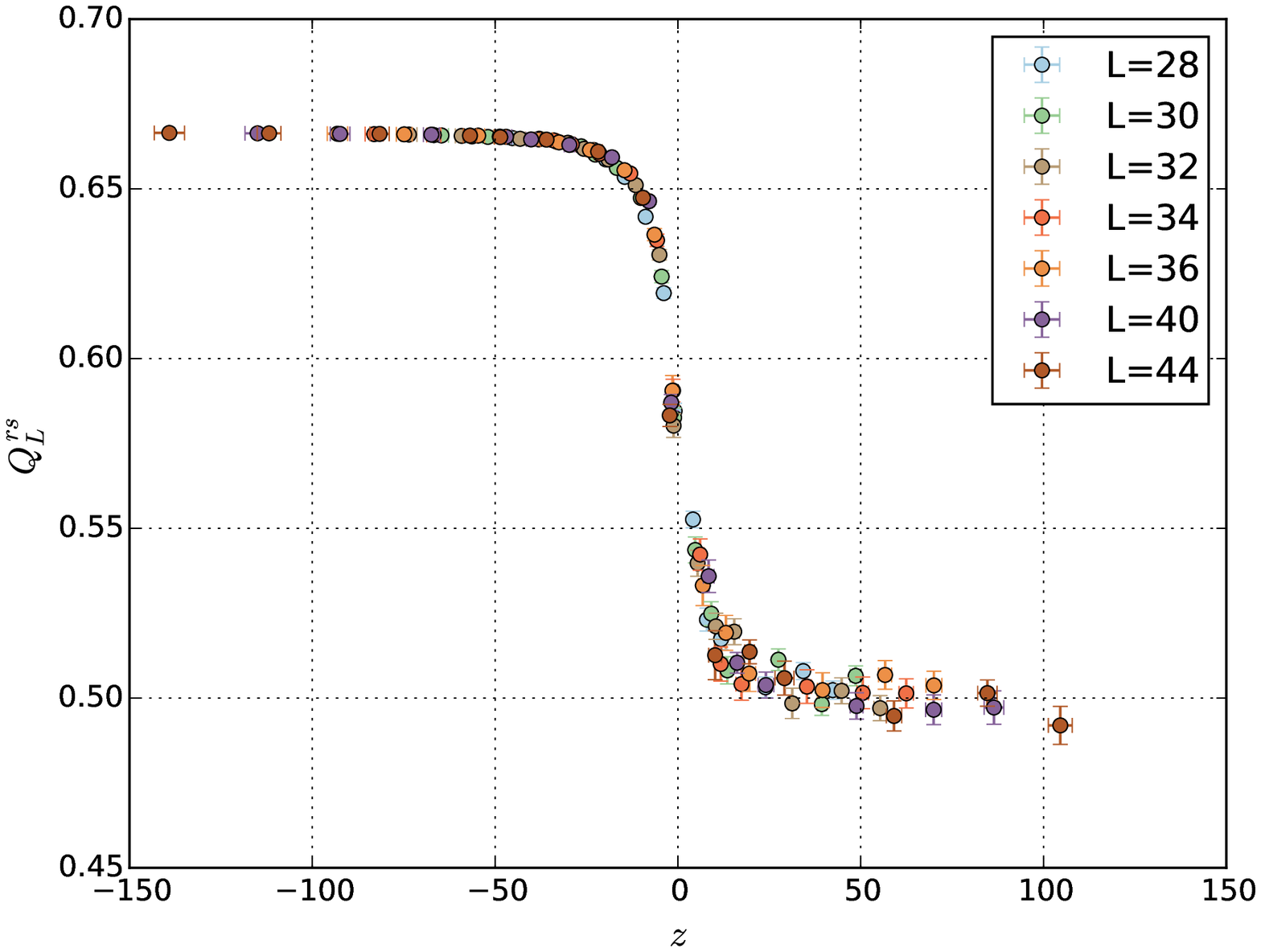}
\\
% \vspace*{0.5 cm}
\hspace*{0.8 cm}(a) \hspace*{7.8 cm}(b)
\caption{
  \label{fig:binfit} 
(a) Fitting of Binder's Cumulant using our analytical formula.  
(b) Plotting data with respect to the scaling variable $z$ using 
the fit parameters we found in (a).
}
\end{center}
\end{figure}
\begin{equation}
\label{bin}
 Q_L = 1-\frac{\langle \varphi^4 \rangle}{3\langle \varphi^2 \rangle^2} 
 = 1-\frac{\bar{\varphi}_7(z)\bar{\varphi}_3(z)}{3\bar{\varphi}_5(z)^2}.
\end{equation}
The result is shown in Fig.~\ref{fig:binfit}a, where the only two fit parameters 
come from the renormalised $\lambda$ at two different phases.  
We only choose data where the renormalised quadratic coupling $\hat{M}^2(m_P) \in [-0.004,0.004]$
to ensure the applicability of leading-order perturbation theory.
The fitting parameters we extract from this procedure are $\lambda(m_P^{bro.}) \equiv \lambda_b  = 0.0831$ 
and $\lambda(m_P^{sym.}) \equiv \lambda_s = 0.0757$.
The $p$-value of this Binder's Cumulant analysis is $p=0.0001$.
In Fig.~\ref{fig:binfit}b
we reconstruct the scaling variable Eq.~(\ref{O4_scaling_variable}) using the value we 
found in Fig.~\ref{fig:binfit}a, and observe a very nice scaling behaviour.

\begin{figure}[!t]
\begin{center}
\vspace*{0.5cm}
\includegraphics[width=0.45\columnwidth]{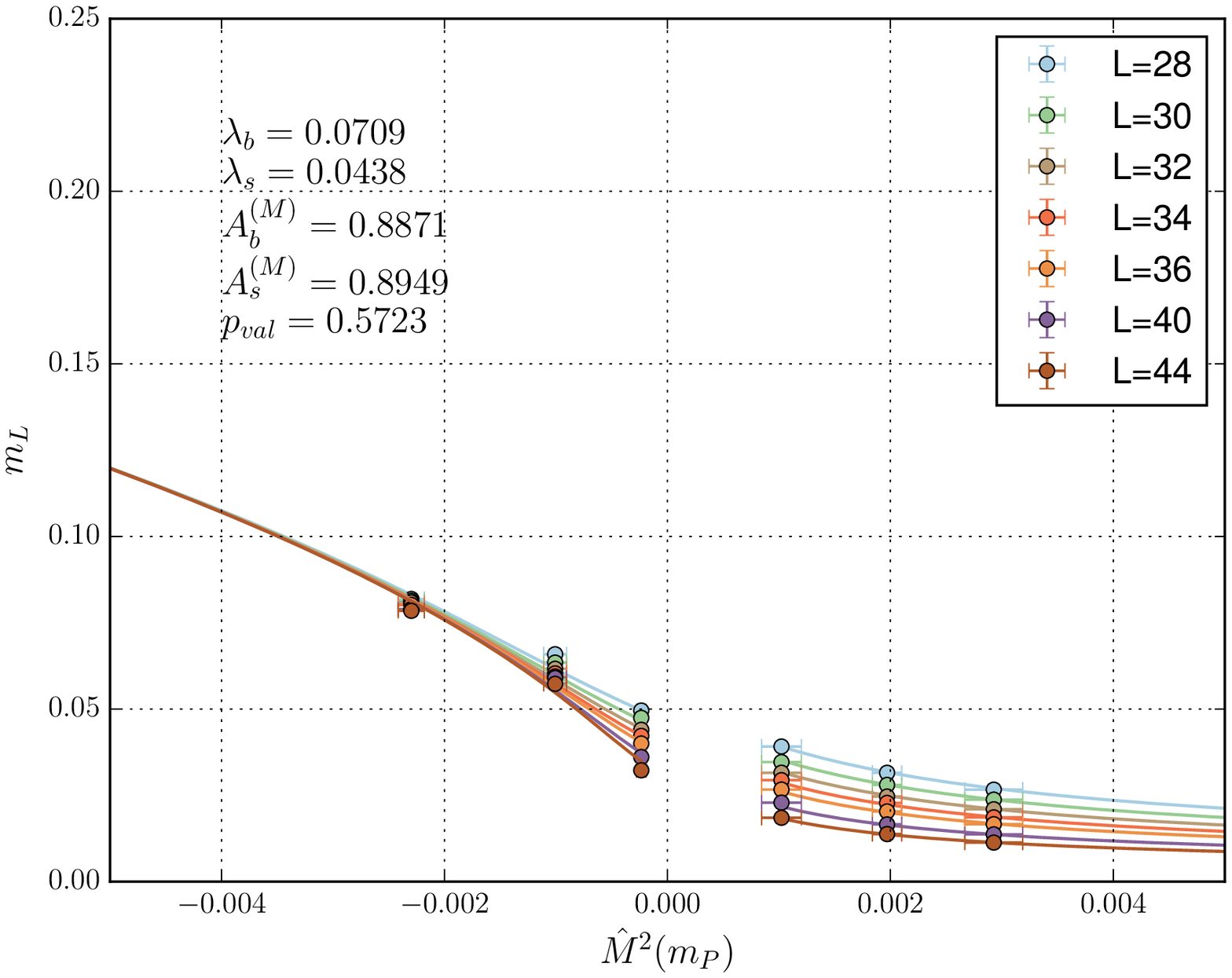}
\hspace*{13 mm}
\includegraphics[width=0.45\columnwidth]{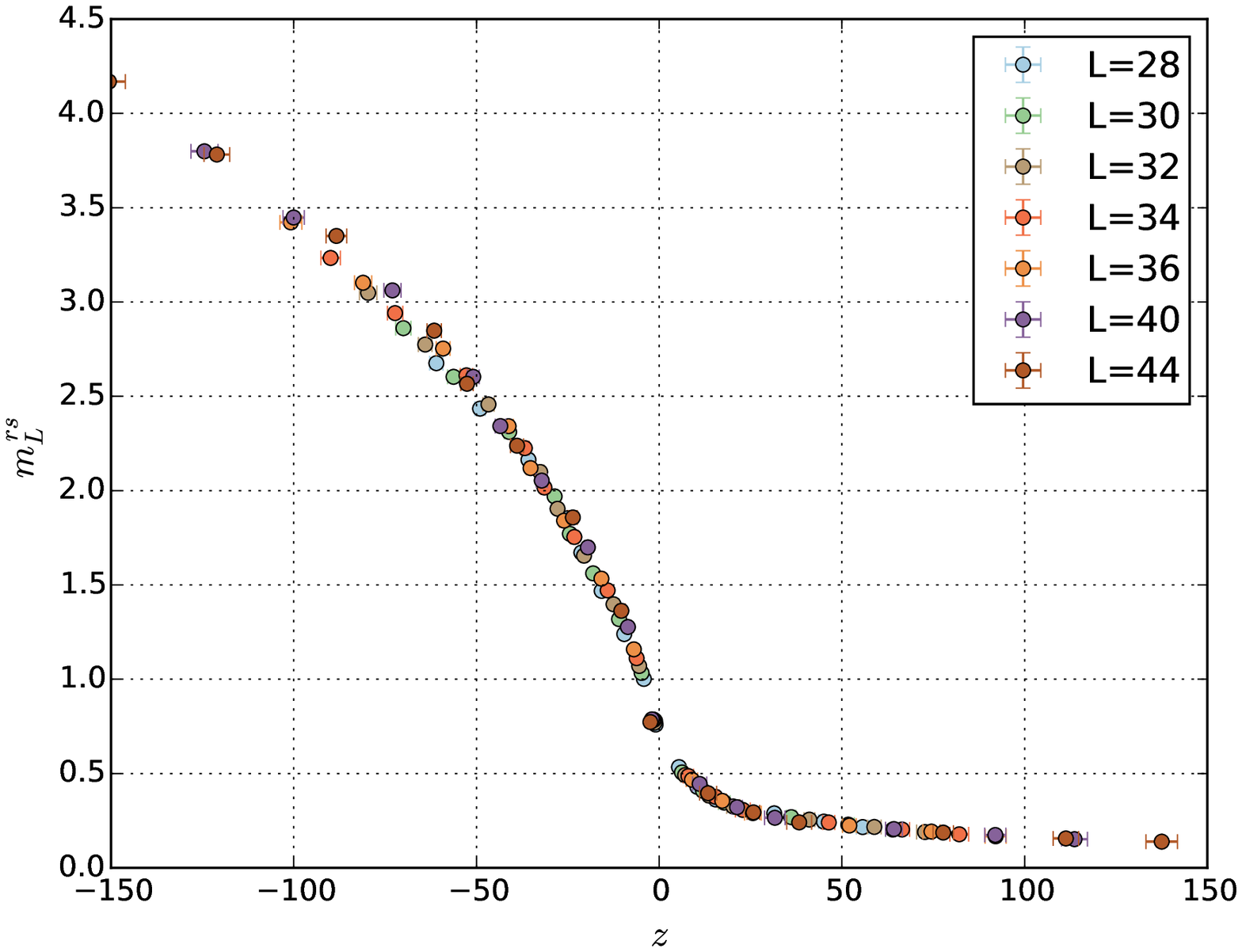}
\\
\hspace*{0.8 cm}(a) \hspace*{7.8 cm}(b)
\\
\vspace*{0.50 cm}
\includegraphics[width=0.45\columnwidth]{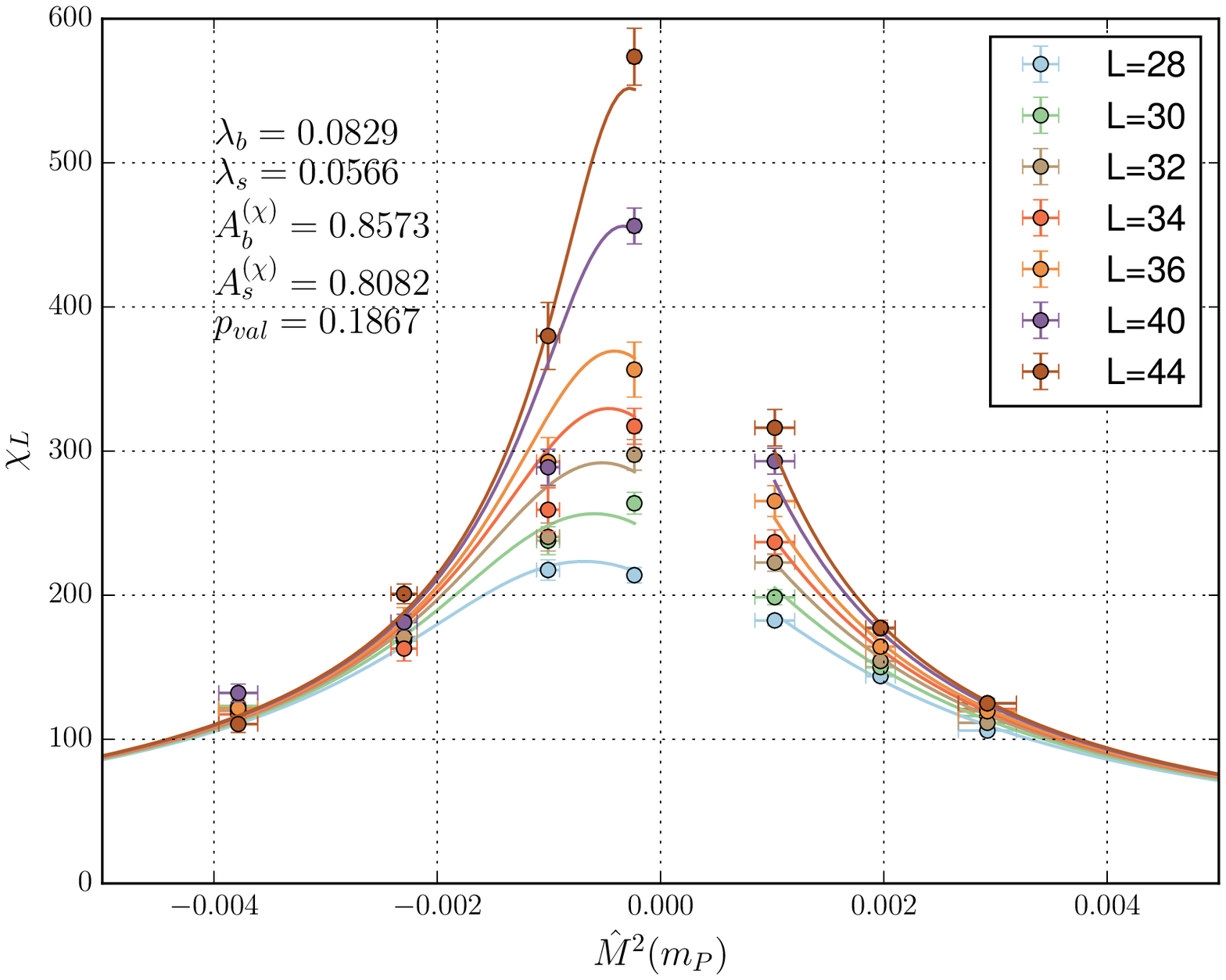}
\hspace*{13 mm}
\includegraphics[width=0.45\columnwidth]{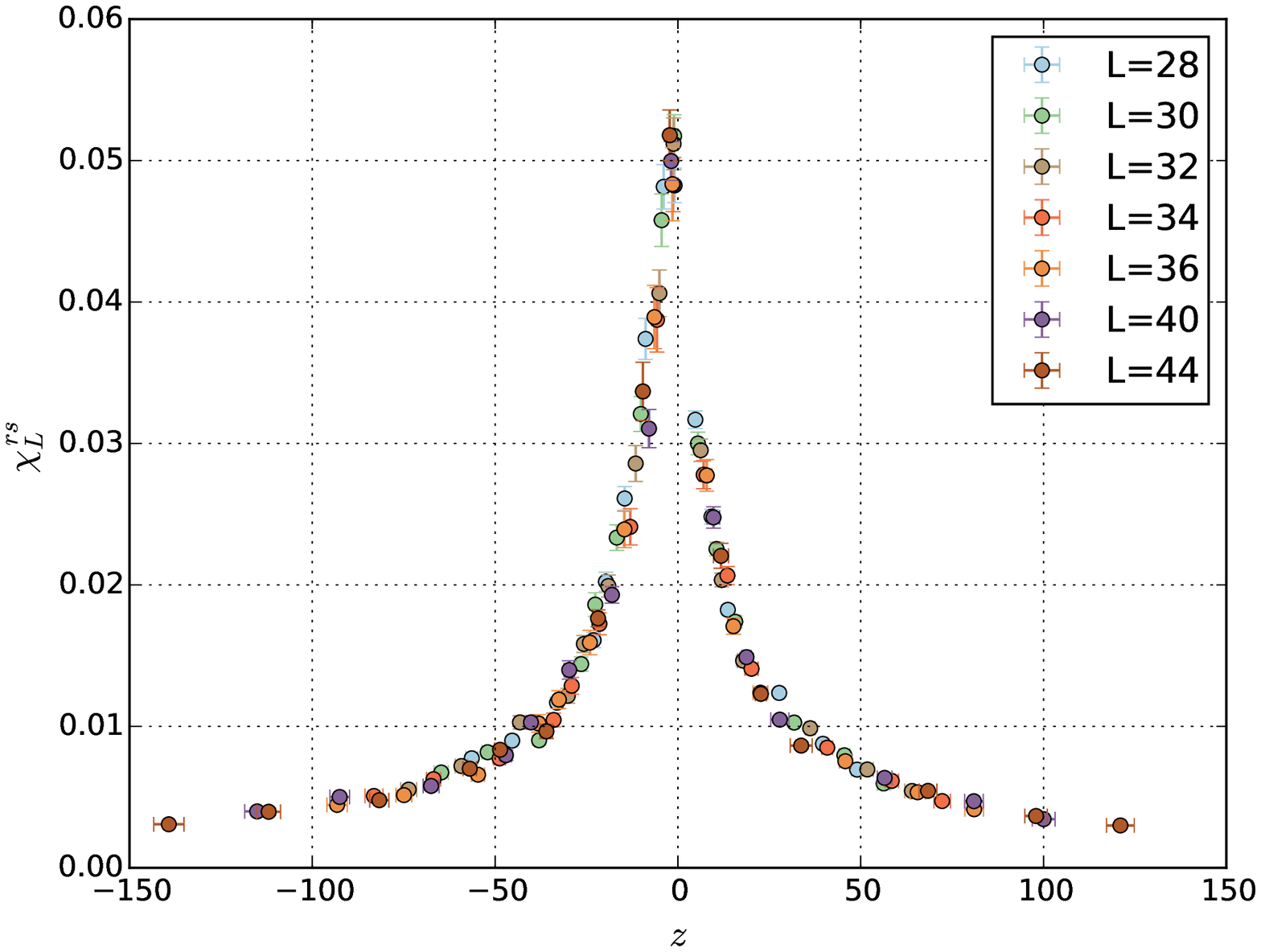}
\\
\hspace*{0.8 cm}(c) \hspace*{7.8 cm}(d)
\caption{
  \label{fig:fitms} 
The fitting to the VEV (a) and the susceptibility (c) with their scaling functions.
(b) Plotting the rescaled VEV, $M^{rs}_L $,
with the scaling variable using fit parameters found in (a).
(d) Plotting the rescaled susceptibility, $\chi^{rs}_L $,
as a function of the scaling variable using fit parameters found in (b).
}
\end{center}
\end{figure}

Similarly, we perform a fit to the VEV, $M_L$, 
and the susceptibility, $\chi_L$.
Their definition and scaling functions are,
\begin{eqnarray}
 \label{mag}
 M_L &=& \langle \hat{\varphi} \rangle
     = 2^{-1/4} A^{(M)} \hat{L}^{-1} \left[ \frac{\lambda(m_P)}{1+\frac{6}{\pi^2}\log(\hat{m}_P\hat{L})} \right]^{-1/4}
     \frac{\bar{\varphi}_4(z)}{\bar{\varphi}_3(z)}, \\
 \label{sus}
 \chi_L &=& 2\hat{L}^4 \left( \langle \hat{\varphi}^2 \rangle -\langle \hat{\varphi} \rangle^2 \right)
     = 2^{1/2} A^{(\chi)} \hat{L}^{2} \left[ \frac{\lambda(m_P)}{1+\frac{6}{\pi^2}\log(\hat{m}_P\hat{L})} \right]^{-1/2}
     \left[ \frac{\bar{\varphi}_5(z)}{\bar{\varphi}_3(z)} - 
     \left( \frac{\bar{\varphi}_4(z)}{\bar{\varphi}_3(z)} \right)^2 \right],
\end{eqnarray}
where the overall factors, $A^{(M)}$ and $A^{(\chi)}$, are the relevant wavefunction 
renormalisation constants.  
The result of fitting the VEV with the scaling function Eq.~(\ref{mag}) is shown in Fig.~\ref{fig:fitms}a. 
Following the conventions for the Binder's Cumulant, we found $\lambda_b = 0.0709$, $\lambda_s = 0.0438$,  
and the prefactors $A^{(M)}(m_P^{(bro.)}) \equiv A^{(M)}_b = 0.8871$, $A^{(M)}(m_P^{(sym.)}) \equiv A^{(M)}_s = 0.8849$ 
and $p = 0.5723$.  
The fit results of the susceptibility, shown in Fig~\ref{fig:fitms}c, give $\lambda_b = 0.0829$, $\lambda_s = 0.0565$, 
$A^{(\chi)}_b = 0.8573$, $A^{(\chi)}_s = 0.8080$ and $p = 0.1867$.  
The universal scaling is also observed as we plot the rescaled VEV, 
$M^{rs}_L = M_L\times \hat{L} \left[ \frac{\lambda(m_P)}{1+\frac{6}{\pi^2}\log(\hat{m}_P\hat{L})} \right]^{1/4}$
, and the rescaled susceptibility, 
$\chi^{rs}_L = \chi_L\times \hat{L}^{-2} \left[ \frac{\lambda(m_P)}{1+\frac{6}{\pi^2}\log(\hat{m}_P\hat{L})} \right]^{1/2}$,
as a function of the scaling variable, as shown in Fig.~\ref{fig:fitms}b and Fig.~\ref{fig:fitms}d.
These fits, in terms of $p$-value, are better 
than the fit to Binder's Cumulant.  At present state we only perform fit to the mean values.  
We will include the full bootstrap error in these fits in the near future.

\section{Summary and Outlook}

In this work we investigate the FSS of HY models in the vicinity of a GFP.  
Employing a proper choice of renormalisation scheme, we are able to 
establish a strategy for studying such systems.
The feasibility test of our strategy is performed with the scalar $O(4)$ model.
However to clearly identify the the logarithmic corrections at this stage is not possible.
The fits to the lattice data with logarithms removed from the scaling functions are almost as
good as the current result.
In the future we plan to have larger lattices, higher statistics, 
data closer to the phase transition, as well as the inclusion of
the effects of the massive mode contributions in the infinite-volume extrapolation.
Eventually we will apply this strategy to explore the HY model at both weak and strong couplings.

\section*{Acknowledgments}

We thank Chung-Wen Kao for valuable discussions.  
The simulations have been performed at HPC facilities at National Chiao-Tung University.  
The works are supported by Taiwanese MOST via grant 102-2112-M-009-002-MY3, the support from the 
DAAD-MOST exchange programme via project number 57054177, 
and the support from DFG through the DFG-project  Mu932/4-4.

%%%%%%%%%%%%%%%%%%%%%%%%%%%%%%%%%%%%%%%%%%%%%%%%%%%%%%%%%%%%%%%%%%%%
%%%                                                              %%%
%%%                         BIBLIOGRAPHY                         %%%
%%%                                                              %%%
%%%%%%%%%%%%%%%%%%%%%%%%%%%%%%%%%%%%%%%%%%%%%%%%%%%%%%%%%%%%%%%%%%%%

\bibliographystyle{apsrev} 
\bibliography{refs.bib} 

\end{document}